\documentstyle[psfig]{lamuphys}
\makeatletter
\let\chapter\hid@chapter
\makeatother
\begin{document}
\pagenumbering{arabic}
\title{Wolf-Rayet and LBV Nebulae as the Result of Variable and
Non-Spherical Stellar Winds}
\titlerunning{WR \& LBV Nebula Dynamics}

\author{Mordecai-Mark Mac Low}

\institute{Max-Planck-Institut f\"ur Astronomie, K\"onigstuhl 17,
D-69117 Heidelberg, Germany}

\maketitle
\begin{abstract}
The physical basis for interpreting observations of nebular morphology
around massive stars in terms of the evolution of the central stars is
reviewed, and examples are discussed, including NGC 6888, OMC-1, and
$\eta$ Carinae.
\end{abstract}

\section{Introduction}

The nebulae observed around massive, post-main sequence stars appear
to consist of material ejected by the central stars during earlier
phases of their evolution, rather than ambient interstellar matter.
Models of these nebulae can be used to constrain the mass-loss history
of the stars, giving an important input for stellar evolution models.
Understanding the structure of these nebulae also clarifies the
initial conditions for the resulting supernova remnants, which will
interact with the circumstellar material for most of their observable
lifetimes before encountering the surrounding interstellar medium.

\begin{figure} 
\psfig{figure=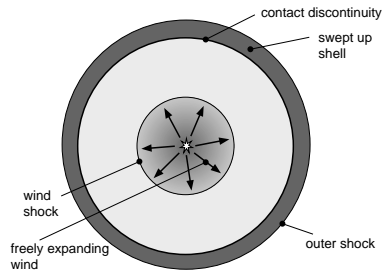,height=1.4in}
\caption{Stellar wind bubble structure.\label{struct}}
\end{figure}

A strong stellar wind 
sweeps the surrounding interstellar gas into a stellar wind bubble as
shown in Figure~\ref{struct}.  The stellar wind expands freely until
it reaches a termination shock. 
If this shock is adiabatic, the hot gas sweeps up the surrounding ISM
into a dense shell, forming a pressure-driven or energy-conserving
bubble that sweeps up the surrounding ISM into a dense shell growing
as $R \propto t^{3/5}$ in a uniform medium (Castor, McCray, \& Weaver
1975).  Should the termination shock be strongly radiative due to high
densities or low velocities in the wind, the bubble only conserves
momentum and will grow as $R \propto t^{1/2}$ (Steigman, Strittmatter,
\& Williams 1975). For more general discussions of blast waves in
non-uniform media, see Ostriker \& McKee (1988), and Bisnovatyi-Kogan
\& Silich (1995), as well as Koo \& McKee (1990).

When these stars leave the main sequence, they pass through phases of
greatly increased mass loss.  These slow, dense winds expand 
into the rarefied interior of the main sequence bubble until their ram
pressure $\rho_w v_w^2$ drops below the pressure of the bubble.
(Should the main sequence bubble have cooled relatively quickly, this
may never occur.)  As the mass loss rate and velocity of the central
wind vary during the post-main sequence evolution of the central star,
these denser winds can in turn be swept up, producing the observed
ring nebulae around evolved massive stars.

During their evolution, these nebulae are subject to a number of
hydrodynamical instabilities,
as well as thermal instabilities ({\em e.\ g.} Strickland \& Blondin
1995).  I explain how understanding the physical basis of the
hydrodynamical instabilities gives insight into the dynamics of
observed nebulae.  High-resolution observations of nebular morphology
can thus be used to constrain the mass-loss history of the central
star.

I then describe how typical stellar evolutionary scenarioes can
generate observed nebular morphologies, and show semi-analytic and
numerical models derived from these scenarioes.  For example, a star
with a stellar wind varying from fast to slow and back again will
have a clumpy circumstellar nebula due to hydrodynamical instabilities
in the shell (Garc\'{\i}a-Segura, Langer \& Mac Low
1996). Nonspherical winds and stellar motion can add to the
morphological richness of the resulting nebulae, as in the nebula
around $\eta$ Car (Langer, Garc\'{\i}a-Segura, \& Mac Low 1998).  A
recent review of this topic is Frank (1998).

\section{Shell Instabilities}

Gas swept up by a stellar wind will usually be subject to different
instabilities. 
An adiabatic, decelerating shell with a density
contrast across the shock of less than 10 in a uniform medium is
stable.  However, relaxing any of these constraints will lead to
instabilities as I now describe.

\subsection{Rayleigh-Taylor Instability}

If the swept-up shell is denser than the stellar wind, as will be true
in virtually all cases where a shell exists at all, the shell will be
subject to RT instabilities if the contact discontinuity
between the shocked stellar wind and the shell accelerates.  This can
be due to an external density gradient steeper than $r^{-2}$ or to a
sufficiently fast increase in the power of the central stellar wind,
though these two mechanisms will lead to different shell morphologies,
as I discuss below.  

\begin{figure}[tb]
\vspace*{-0.32in}
\psfig{figure=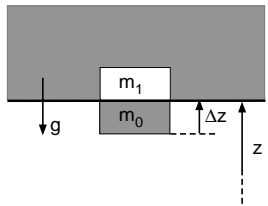,width=2.5in,rheight=1.2in}
\caption{Rayleigh-Taylor instability\label{rt}}
\end{figure}

The RT instability occurs when the effective gravity due
to acceleration points from a denser to a more rarefied gas.  We can
understand its driving mechanism by considering how the potential
energy will change if we interchange a parcel of dense gas having mass
$m_1$ with a parcel of more rarefied gas having $m_0 < m_1$, as shown
in Figure~\ref{rt}.  The potential energy before the interchange is
given by $E_i =  m_1 g z + m_0 g (z - \Delta z)$ which is greater than
the potential energy after the interchange $E_f = m_0 g z + m_1 g (z -
\Delta z)$ due to the difference in the masses.  The decrease in
potential energy drives an exponentially growing interchange of the
two fluids.

\begin{figure}[tb]
\psfig{figure=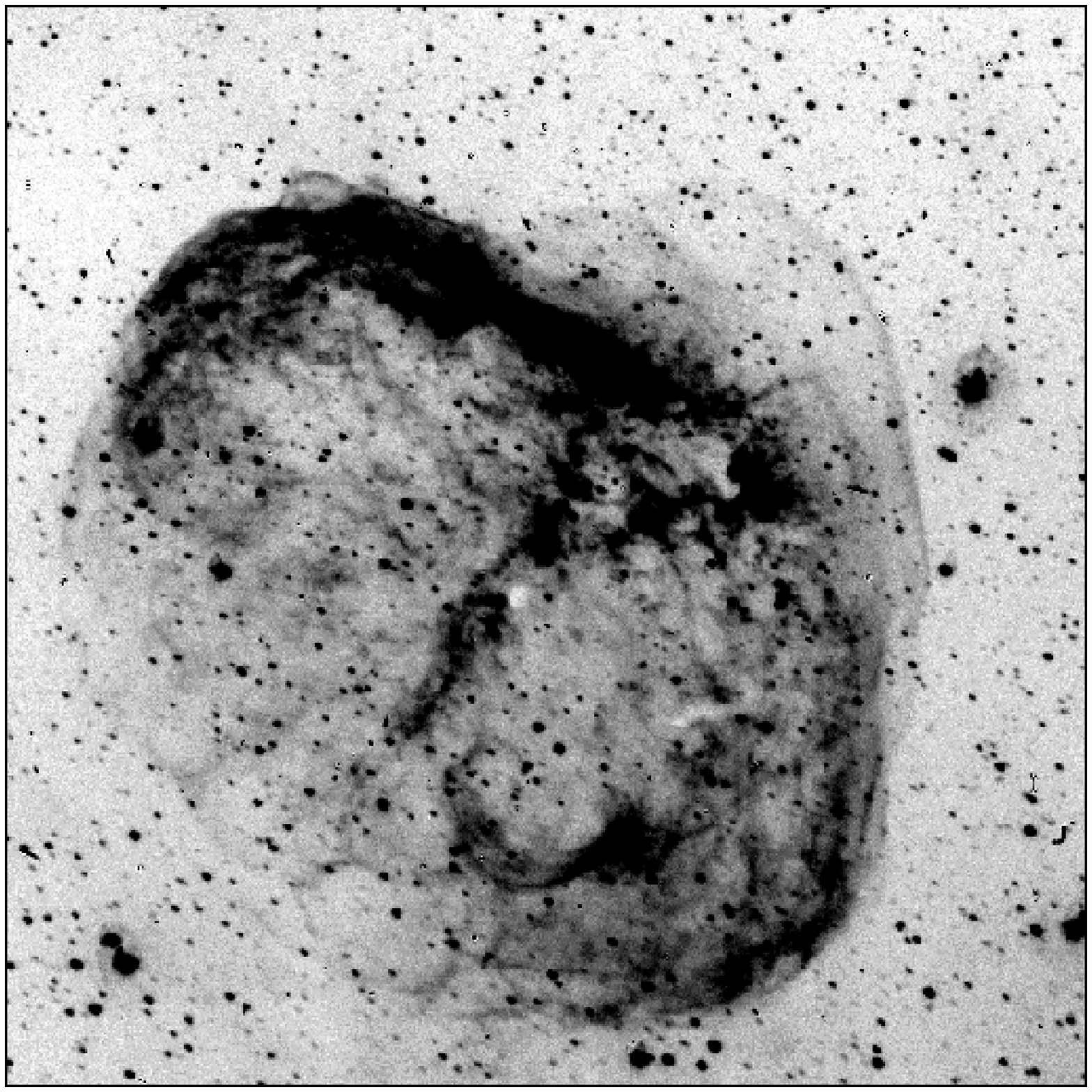,width=2.25in}
\vspace*{-2.25in} \hspace*{2.25in}
\psfig{figure=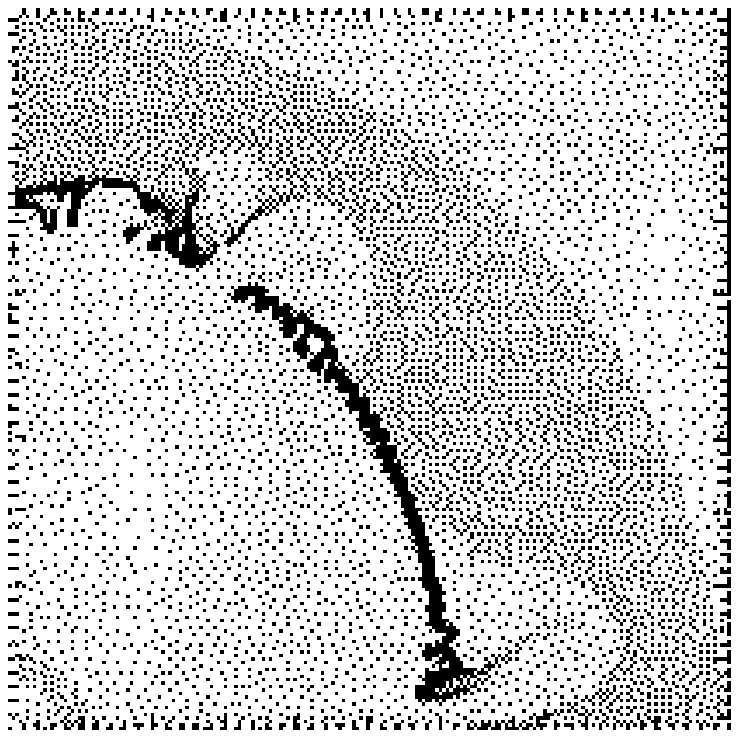,width=2.25in}
\caption{Comparison between morphology of NGC~6888 in O~[{\sc iii}]
(in an image taken by K. B. Kwitter with the Burrell-Schmidt telescope
of the Warner and Swasey Observatory, Case Western Reserve
University), and a numerical simulation of RT instability due to a
fast wind sweeping over the end of a slow, dense wind
(Garc\'{\i}a-Segura \& Mac Low 1995).  The model image shows a
cross-section of the density structure in grayscale with black
indicating high density and white low density. \label{rtmorph}}
\end{figure}

When a RT instability occurs due to an external density
gradient, dense fragments of shell are left behind as the less dense
interior expands out beyond them, creating the characteristic bubble
and spike morphology seen, for example, in models of superbubble
blowout (e.\ g.\ Mac Low, McCray, \& Norman 1989).  The Wolf-Rayet
ring nebula NGC~6888 shown in Figure~\ref{rtmorph} provides another
example.  Here a fast, rarefied Wolf-Rayet wind has swept up the slow,
dense red supergiant wind that preceded it.  While it was still
sweeping up the slow wind, it was 
marginally stable to RT
instabilities.  However, at the outer edge of the slow wind, the
sharp density
gradient triggers RT instabilities, as modelled by 
Garc\'{\i}a-Segura \& Mac Low (1995) with the astrophysical gas
dynamics and magnetohydrodynamics code ZEUS\footnote{Available by
registration with the Laboratory for Computational Astrophysics at
lca@ncsa.uiuc.edu} (Stone \& Norman 1992).

On the other hand, when a RT instability occurs due to an
increase in power of the driving wind, some of the dense fragments of
shell actually get shot out ahead of the bulk of the fragmenting
shell, producing a markedly different morphology (Stone, Xu, \& Mundy
1995).  Although these fragments represent only a small fraction of
the total mass of the shell, they can produce a very striking set of
bow shocks in their wake.  An example of this occurring around one or
more pre-main sequence stars is given by the ``bullets'' observed
around OMC-1 (Lane 1989, Allen \& Burton 1993) in the Orion star
forming region, as confirmed by McCaughrean \& Mac Low (1997).

\subsection{Vishniac Instability}

If a pressure-driven shell is decelerating, but thin, with a
density contrast across the shock of at least 25 for a stellar-wind
bubble expanding into a uniform medium (Ryu \& Vishniac 1988), or 10
for a point explosion (Ryu \& Vishniac 1987), it will be subject to the
Vishniac overstability (Vishniac 1983).  This has been confirmed
experimentally using blast waves generated by high-powered lasers
propagating into gases with low and high adiabatic indexes (Grun et
al.\ 1991).

\begin{figure}[tb]
\vspace*{-0.6in}
\psfig{figure=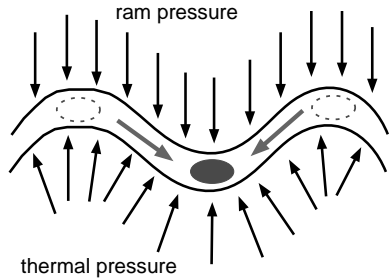,height=2in,rheight=2in}
\caption{Vishniac instability mechanism\label{vish}}
\end{figure}

The mechanism of the Vishniac overstability can be understood by
considering a thin, decelerating shell driven from within by a
high-pressure region, as shown in Figure~\ref{vish}.  From within it
is confined by thermal pressure acting normal to the shell surface, as
adjacent regions can communicate with each other by sound waves, while
from outside it is confined by ram pressure acting parallel to the
velocity of propagation, as the shell moves supersonically into the
surrounding gas.  In equilibrium, these two forces remain in balance.
Should the shell be perturbed, however, the thermal pressure will
continue to act normally, but the ram pressure will now act obliquely,
giving a transverse resultant force that drives material from
``peaks'' into ``valleys'' of the shell.  The denser valleys will be
decelerated less than the rarefied peaks, however, so that the
positions of peaks and valleys are interchanged after some time.
Vishniac (1983) showed that this overstable oscillation can grow as
fast as $t^{1/2}$.  It saturates when the transverse flows in the
shell become supersonic and form transverse shocks, so that the end
result of Vishniac instability is a shell with transonic turbulence
and moderate perturbations (Mac Low \& Norman 1993).

\subsection{Nonlinear Thin Shell Instability}

Should the driving wind cool immediately behind its termination shock,
for example because of exceptionally high mass-loss rates, it can form
a decelerating shell that is momentum-driven rather than
pressure-driven, so that it is effectively confined on both sides by
ram pressure from shocks.  Such a shell is not subject to the Vishniac
instability, and is, in fact, linearly stable.  However Vishniac
(1994) has shown that if the shell is strongly perturbed, it will
still be subject to a nonlinear thin shell instability (NTSI).  When
the shocks are oblique enough to the direction of flow, they will bend
the streamlines passing through them, so that mass is driven towards
the extrema of the perturbation.  Numerical simulations by Blondin \&
Marks (1996), using a piecewise parabolic hydrocode called VH-1, have
shown that the end result is a catastrophic breakup of the shell into
a turbulent layer that grows in time.

\section{A Final Example:  Eta Carinae}

As an example of how knowledge of these different instabilities can be
used to constrain the evolution of a star, consider the example of the
Homunculus Nebula around $\eta$ Car.  Langer, Garc\'{\i}a-Segura,
\& Mac Low (1998) computed several two-dimensional models of it using
ZEUS, following a basic scenario in which a luminous blue star with a
fast stellar wind undergoes an outburst during which it has a much
slower and denser wind strongly shaped by rotation, as described by
Bjorkman \& Cassinelli (1993), but then reverts to its previous state
with a fast, rarefied wind.  
\begin{figure}[t]
\vspace*{-1.25in}\hspace*{-0.5in}
\psfig{figure=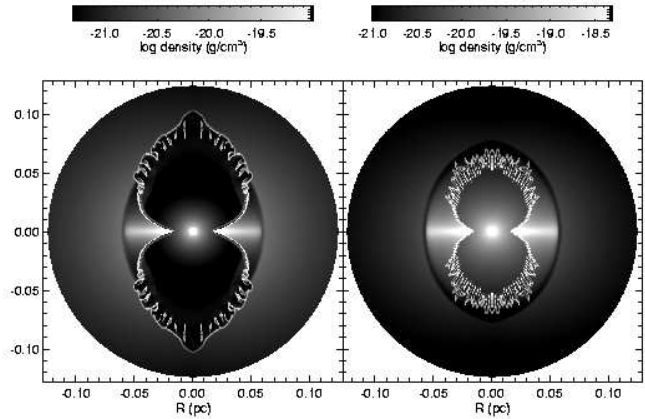,rheight=3.2in,height=4in,angle=90}
\caption{Two-dimensional density distributions from models of $\eta$
Car by Langer, Garc\'{\i}a-Segura, \& Mac Low (1998) with faster
and slower post-outburst wind showing Vishniac instabilities and the
NTSI respectively.  Note how the faster wind model resembles a
cross-section through the cauliflower-like observed
lobes. \label{eta}}
\end{figure}
They chose two different values for the
post-outburst wind, one consistent with current observed values of
$\dot{M} = 1.3 \times 10^{-3} M_{\sun} \mbox{ yr}^{-1}$ and $v_w =
1300 \mbox{ km s}^{-1}$, and one with a faster, lower mass loss wind
having $\dot{M} = 1.7 \times 10^{-4} M_{\sun} \mbox{ yr}^{-1}$ and
$v_w = 1800 \mbox{ km s}^{-1}$.  As shown in Figure~\ref{eta}, the
slower, denser wind cools upon shocking, forming a momentum-driven
shell that fragments due to the NTSI, producing a sharp, spiky shell
morphology.  On the other hand, the faster wind does not cool
completely, and forms a bubble subject to Vishniac instabilities,
giving it a much more curved, cauliflower-like appearance.  Comparison
to the high-resolution observations (Humphreys \& Davidson
1994; Morse, Davidson, \& Ebbets 1997) reveals that the actual
morphology strongly resembles a three-dimensional version of the
model with the faster wind.  Langer et al.\ (1998) suggest that
this reflects the typical behavior of the wind over the century since
the outburst, and that the current wind properties are actually
exceptional, and perhaps even indicative of another outburst on its
way.  This suggestion is supported by the gradual
brightening of $\eta$ Car over the last decades (Humphreys \& Davidson 1994).

This work has made use of the NASA Astrophysical Data
System Abstract Service.  I thank the organizers for their invitation
and their support of my attendence at this meeting.

\end{document}